\documentclass[prd,aps,preprintnumbers,twocolumn,showpacs,nofootinbib,superscriptaddress,notitlepage,floatfix]{revtex4-1}%,
\usepackage{amsthm,amsmath,amssymb}
\usepackage{xcolor}
\usepackage{graphicx}
\usepackage{multirow}
\usepackage[colorlinks, citecolor=blue,anchorcolor=blue,menucolor=blue, linkcolor=blue,filecolor=blue,runcolor=blue,urlcolor=blue,frenchlinks=true]{hyperref}
\usepackage{multirow}

\begin{document}
\title{The Gluon Moment and Parton Distribution Function
of the Pion from \texorpdfstring{\boldmath{$N_f=2+1+1$}}{Nf=2+1+1} Lattice QCD }

\author{William Good}
\affiliation{Department of Physics and Astronomy, Michigan State University, East Lansing, MI 48824}
\author{Kinza Hasan}
\affiliation{Department of Physics and Astronomy, Michigan State University, East Lansing, MI 48824}
\author{Allison Chevis}
\affiliation{Department of Physics and Astronomy, Michigan State University, East Lansing, MI 48824}
\author{Huey-Wen Lin}
\email{hwlin@pa.msu.edu}
\affiliation{Department of Physics and Astronomy, Michigan State University, East Lansing, MI 48824}

%%%%%%%%%%%%%%%%%%%%%%%%%%%%%%%%%%%%%%%%%%%%%%%%%%%%%%%%%%%%%%%%%%%%%%%%%%%%%%%%
\preprint{MSUHEP-23-027}


%%%%%%%%%%%%%%%%%%%%%%%%%%%%%%%%%%%%%%%%%%%%%%%%%%%%%%%%%%%%%%%%%%%%%%%%%%%%%%%%
\begin{abstract} 
We present the first calculation of the pion gluon moment from lattice QCD in the continuum-physical limit.
The calculation is done using clover fermions for the valence action with three pion masses, 220, 310 and 690~MeV, and three lattice spacings, 0.09, 0.12, and 0.15~fm, using ensembles generated by MILC Collaboration with 2+1+1 flavors of highly improved staggered quarks (HISQ).
On the lattice, we nonperturbatively renormalize the gluon operator in RI/MOM scheme using the cluster-decomposition error reduction (CDER) technique to enhance the signal-to-noise ratio of the renormalization constant.
We extrapolate the pion gluon moment to the continuum-physical limit and obtain $0.364(38)_{\text{stat}+\text{NPR}}(36)_\text{mixing}$ in the $\overline{\text{MS}}$ scheme at 2~GeV, with first error being the statistical error and uncertainties in nonperturbative renormalization, and the second being a systematic uncertainty estimating the effect of ignoring quark mixing.
Our pion gluon momentum fraction has a central value lower than two recent single-ensemble lattice-QCD results near physical pion mass but is consistent with the recent global fits by JAM and xFitter and with most QCD-model estimates. 
\end{abstract}

\maketitle

%%%%%%%%%%%%%%%%%%%%%%%%%%%%%%%%%%%%%%%%%%%%%%%%%%%%%%%%%%%%%%%%%%%%%%%%%%%%%%%%
\section{Introduction}

The lightest known hadron of quantum chromodynamics (QCD), the pion is the Nambu-Goldstone boson associated with dynamical chiral symmetry breaking, critical to furthering our understanding of the emergence of physical mass.
Theoretical and experimental study of pion structure is a necessary precursor to answering some of the highlight science questions in current QCD research;
we refer readers to recent reviews in Refs.~\cite{Roberts:2021nhw,Arrington:2021biu,Aguilar:2019teb}.
Better discerning the structure of the pion requires that we increase our knowledge of the pion parton distribution function (PDF), especially its gluonic content.
The determination of the pion PDF from experimental data~\cite{Badier:1983mj,Betev:1985pf,Conway:1989fs,Wijesooriya:2005ir,Aicher:2010cb} beyond its valence quarks is limited in scope.
Since the pion's decay makes it inaccessible as a scattering target, its gluon and sea-quark content remains less well constrained empirically in comparison to nucleons. 
Existing analyses of the pion PDFs primarily utilize Drell-Yan data from CERN and Fermilab with additional constraints determined from sources such as leading-neutron electroproduction data from HERA~\cite{Barry:2018ort,Cao:2021aci}.
A host of planned facilities, such as Brookhaven National Laboratory's Electron-Ion Collider (EIC)~\cite{Achenbach:2023pba}, the Electron-Ion Collider in China (EicC)~\cite{Anderle:2021wcy}, and CERN's COMPASS++ and AMBER experiments~\cite{Adams:2018pwt}, anticipate highly advanced measurement capabilities, probing energy regimes which will facilitate experiments uniquely useful for studying light pseudoscalar mesons, especially the gluon structure of pion.

Lattice QCD (LQCD) provides a first-principles approach to studying the pion PDF in a nonperturbative context.
The ``quasi-PDF'' approach, also called Large-Momentum Effective Theory (LaMET), allows the $x$ dependence of PDFs to be studied using LQCD~\cite{Ji:2014gla,Ji:2017rah,Ji:2013dva,Zhang:2018diq,Wang:2019tgg} and has been broadly used to study pion valence-quark parton distributions~\cite{Lin:2023gxz,Gao:2022iex,Gao:2020ito,Zhang:2018nsy,Izubuchi:2019lyk,HadStruc:2022yaw,Balitsky:2021qsr}.
LaMET shows that the matrix elements of a time-dependent lightcone operator can be extracted by utilizing large-momentum expansion of quasi-operators in a hadron state having large momentum. 
Another commonly used approach, the ``pseudo-PDF'' method~\cite{Orginos:2017kos,Radyushkin:2018cvn,Balitsky:2019krf,Balitsky:2021cwr}, has been used to access the gluon PDFs $g(x)$ of the nucleon, pion, and kaon~\cite{Fan:2022kcb,Salas-Chavira:2021wui,Fan:2021bcr,Fan:2020cpa,Fan:2018dxu,Khan:2022vot,HadStruc:2021wmh,Delmar:2023agv}. 
However, these methods give the ratio $g(x)/ \langle x_g \rangle$ with $\langle x_g \rangle =  \int_{0}^{1} dx \, x g(x)$.
The gluon moment also plays an important role in determining the momentum, spin, and mass decompositions of hadrons, which are important topics of QCD research. 
The PDF computed via quasi-PDF or pseudo-PDF approach can in principle be used to derive the gluon moment associated with a particular distribution.
However, considering the limiting aspects of this approach, such as systematic uncertainties at large- and small-$x$, there is a definite motivation to explore the gluon moment via alternative methods.

\begin{table*}[htbp!]
\centering
\begin{tabular}{|c|c|c|c|c|c|c|c|c|}
\hline
  Group & $N_f$ &  $a$ (fm) & $M_\pi^\text{val}$ (MeV) & Fermion & $N_\text{meas}$  &  Renorm. &  G-smearing & $\langle x \rangle_g $\\
\hline
\hline
 MIT18~\cite{Shanahan:2018pib} & 2+1 &  0.12 &  $450$ &  clover  & 572,663 & RI-MOM & Wilson flow & $0.61(9)_\text{stat}$\\
 \hline
 ETMC21~\cite{ExtendedTwistedMass:2021rdx} & 2+1+1 & 0.08 & $139.3$ & TM & 149,000 & RI$'$-MOM & 10-stout &  $0.52(11)_\text{stat}(^{+02})_\text{mixing}$
 \\
\hline
MIT23~\cite{Hackett:2023nkr} &  2+1 & 0.09 & 170 & clover & 2,571,264 & RI-MOM & Wilson flow & $0.546(18)_\text{stat}$ \\
\hline
MSULat23 & 2+1+1 &  $[0.09,0.15]$  &
$[220,700]$\footnote{clover-on-HISQ mixed action with valence pion masses tuned to lightest sea-quark ones} & clover & $10^5$--$10^6$ & RI-MOM &
 5-HYP &
$0.364(38)_{\text{stat}+\text{NPR}}(36)_\text{mixing}$\\
(this work) & & & & & & & &  \\
\hline
\end{tabular}
\caption{
Summary of dynamical lattice calculations of the nucleon gluon moment sorted by year.
The columns from left to right show for each calculation:
the number of flavors of quarks in the QCD vacuum ($N_f$),
the lattice spacing ($a$) in fm,
the valence pion mass ($M_\pi^\text{val}$) in MeV,
the valence fermion action (``Fermion''), where ``TM'' stands for twisted-mass fermion action,
the number of measurements of the nucleon correlators ($N_\text{meas}$),
the renormalization method (``Renorm.'') indicating 1-loop perturbative calculations or RI-MOM nonperturbative renormalization,
the smearing technique used to improve the gluon signals (``G-smearing''),
and the obtained gluon momentum fraction ($\langle x \rangle_g$) renormalized at 2-GeV scale in $\overline{\text{MS}}$ scheme. 
The lattice errors coming from different sources are marked as
``stat'' for statistical, 
``NPR'' for nonperturbative renormalization, and ``mixing'' for the mixing with the quark sector.
Note that all the prior lattice works only study the gluon moment at a single lattice spacing;
this work (labeled as ``MSULat23'') is the only one whose statistical error that has lattice-discretization error included in it.
\label{tab:latticemoments}
}
\end{table*}

There are only a handful of LQCD studies exploring the gluon moment of the pion, using a variety of LQCD configurations.
An early calculation using quenched QCD predicted $\langle x \rangle_g = 0.37(8)(12)$ in 2007, using Wilson fermion action with a lattice spacing $a=0.093$~fm at large pion mass $[600,1100]$~MeV~\cite{Meyer:2007tm}.
Then in 2018, a study by researchers from the MIT group removed the quenched approximation and reported $\langle x_g \rangle = 0.61(9)$ using an $N_f = 2 + 1$ ensemble with $a = 0.1167(16)$~fm and $L^3 \times T = 32^3 \times 96$~\cite{Shanahan:2018pib}. 
Later, the Extended Twisted Mass Collaboration (ETMC) utilized an ensemble with $N_f = 2 + 1 + 1$ dynamical twisted-mass fermions with clover term, a lattice volume $L^3 \times T = 64^3 \times 128$, and a lattice spacing $a = 0.08029(41)$~fm~\cite{ExtendedTwistedMass:2021rdx}.
Their $\overline{\text{MS}}$-scheme 2-GeV result for the gluon momentum fraction of the pion given these parameters is $\langle x_g \rangle = 0.52(11)$.  
Recently, in 2023, MIT published a follow-up work, reporting a much improved gluon moment $\langle x_g \rangle = 0.55(2)$ using a finer lattice $a = 0.091(1)$~fm with nearly physical pion mass 170~MeV~\cite{Hackett:2023nkr}. 
Throughout these prior lattice efforts, all the calculations have been done at only a single lattice spacing;
no attempts to remove the discretization effects on the pion moment have been made.
When computing partonic properties of interest using a discretized spacetime, the nonzero lattice spacing must necessarily be accounted for in order to identify the continuum estimate.
We, therefore, present this study of the gluon moment using multiple lattice spacings and extrapolated to the continuum limit.

In this work, we use four ensembles generated with 2+1+1 flavors of highly improved staggered quarks (HISQ) by the MILC collaboration~\cite{MILC:2012znn}.
We construct pion two- and three-point correlation functions using Wilson-clover valence fermions.
In Sec.~\ref{sec:lattice-details}, we present the analysis of the data generated using this clover-on-HISQ formulation, including a study of excited-state contributions in the extraction of ground-state matrix elements.
We use a simultaneous chiral-continuum fit to obtain results at the physical point, which throughout the paper will be defined as taking the continuum limit ($a \to 0$) and physical light-quark masses, as discussed in Sec.~\ref{sec:Results}.
Our final conclusions are presented in Sec.~\ref{sec:Conclusion}.

%%%%%%%%%%%%%%%%%%%%%%%%%%%%%%%%%%%%%%%%%%%%%%%%%%%%%%%%%%%%%%%%%%%%%%%%%%%%%%%%
\section{Bare Lattice Matrix Elements for the Pion Gluon Moment}\label{sec:lattice-details} 

In this paper we present results on the lowest moment of the pion gluon distribution from high-statistics calculations done on four ensembles generated using 2+1+1 flavors of HISQ~\cite{Follana:2006rc} by the MILC Collaboration~\cite{MILC:2012znn}.
The data at three lattice spacings $a$ and three pion masses $M_\pi$ allow us to carry out a simultaneous fit to the physical limit.
We use the same valence-quark parameters for the Wilson-clover fermions as PNDME collaboration; see details in Table~II of Ref.~\cite{Gupta:2018qil}.
The Sheikholeslami-Wohlert coefficient used in the clover action is fixed to its tree-level value with tadpole improvement, $c_\text{sw} = 1/u_0$, where $u_0$ is the fourth root of the plaquette expectation value calculated on the hypercubic (HYP) smeared~\cite{Hasenfratz:2001hp} HISQ lattices.
The mass parameters of light and strange clover quarks are tuned so that the clover-on-HISQ pion masses $M^\text{val}_\pi$ match the HISQ-on-HISQ Goldstone ones composed from sea light and strange quarks, respectively.
For the remainder of this paper, we drop the ``val'' superscript and denote the clover-on-HISQ pion mass $M_\pi$.
The number of measurements made on each ensemble is given in Table~\ref{tab:LatticeParameters}.

\begin{table*}[!htbp]
\centering
\begin{tabular}{|c|c|c|c|c|c|}
\hline
  ensemble & a09m310 & a12m220 & a12m310 (310 MeV) & a12m310 (690 MeV) & a15m310 \\
\hline
  $a$ (fm) & $0.0888(8)$ & $0.1184(10)$ & $0.1207(11)$ & $0.1207(11)$  & $0.1510(20)$ \\
\hline
  $L^3\times T$ & $32^3\times 96$ & $32^3\times 64$ & $24^3\times 64$ & $24^3\times 64$ & $16^3\times 48$ \\
\hline
  $M_{\pi}^\text{val}$ (MeV) & $313.1(13)$ & $226.6(3)$ & $309.0(11)$ & $687.3(6)$ & $319.1(31)$\\
\hline 
  $P_z$ (GeV)  & $[0,1.75]$ & $[0,1.64]$ & $[0,1.71]$  &  $[0,1.71]$  & $[0,1.54]$ \\
\hline
  $N_\text{cfg}$ & 1009 & 957 & 1013 & 1013 & 900   \\
\hline
  $N_\text{meas}$ & $\{387,456\}$ &  1,466,944  & 324,160 & 324,160 & 259,200  \\ 
\hline
  $t_\text{sep}$ & $[7,11]$& $[5,9]$   & $[5,9]$ & $[4,8]$ & $[4,8]$ \\
\hline
\end{tabular}
\caption{
The 2+1+1-flavor HISQ ensembles generated by the MILC collaboration and analyzed in this study with valence pion mass tuned to be as close as possible to the Goldstone sea HISQ pion mass.
The lattice spacing $a$, valence pion mass $M_\pi^\text{val}$ and  
lattice size $L^3\times T$, number of configurations $N_\text{cfg}$, number of total two-point correlator measurements $N_\text{meas}^\text{2pt}$, and source-sink separation $t_\text{sep}$ used in the three-point correlator fits are detailed in this table.
}
\label{tab:LatticeParameters}
\end{table*}

The two-point correlator for a meson $\pi$ calculated on the lattice is
\begin{equation} \label{eq:2pt correlator}
    C_\pi^{\text{2pt}}(P_z;t) = \int d^3y\,e^{-iy_zP_z} \left\langle\chi_{\pi}(\vec{y},t) | \chi_{\pi}(\vec{0},0)\right\rangle ,
\end{equation}
where $P_z$ is the meson momentum in the spatial $z$-direction,
$t$ is the lattice euclidean time,
and $\chi_{\pi}=\bar{q_1}\gamma_{5}q_2$ is the pseudo scalar-meson interpolation operator. 
The quark fields are momentum-smeared via $q(x)+\alpha \sum_j U_j(x) e^{i\frac{2\pi}{L}k\hat{e}_j} q(x+\hat{e}_j)$.
The three-point correlator is calculated by combining the gluon loop with the meson two-point correlator.
We can use it to obtain the matrix elements needed to extract the meson gluon moment.
The three point correlator is
\begin{multline} \label{eq:3pt correlator}
    C_\pi^{\text{3pt}}(P_z;t_{\text{sep}},t) = \\
    \int d^3y\,e^{-iy_zp_z} \left\langle \chi_{\pi}(\vec{y},t_{\text{sep}})|O_{g,tt}(t)|\chi_{\pi}(\vec{0},0) \right\rangle
\end{multline}
where $t_{\text{sep}}$ is the source-sink time separation,
and $t$ is the gluon operator insertion time. 
The operator for the gluon moment $O_{g,tt}(t)$ is
\begin{equation} \label{eq:gluon_op}
    O_{g,\mu \nu} \equiv \sum_{i=x,y,z,t} F^{\mu i}F^{\nu i} -
    \frac{1}{4}\sum_{i,j=x,y,z,t} F^{\mu j}F^{\nu j},
\end{equation}
where the field tensor
\begin{equation} \label{eq:field tensor}
    F_{\mu \nu} = \frac{i}{8a^2g}(\mathcal{P}_{[\mu,\nu]}+\mathcal{P}_{[\nu,-\mu]}+\mathcal{P}_{[-\mu,-\nu]}+\mathcal{P}_{[-\nu,\mu]})
\end{equation}
with the plaquette $\mathcal{P}_ {\mu \nu} = U_{\mu}(x) U_\nu(x+a\hat{\mu}) U_{\mu}\textsuperscript{\textdagger}(x+a\hat{\nu}) U_{\nu}\textsuperscript{\textdagger}(x)$ and  $\mathcal{P}_{[\mu,\nu]} = \mathcal{P}_{\mu,\nu} -\mathcal{P}_{\nu,\mu}$.
The same gluon operator was also used in the calculations of the gluon moment fraction by ETMC and MIT lattice collaborations~\cite{Shanahan:2018pib,ExtendedTwistedMass:2021rdx,Hackett:2023nkr}.
%In order to calculate the gluon momentum fraction from the extracted ground-state matrix elements, we can use 
These correlators have been neglected previously due their worse signal-to-noise ratio relative to those obtained from $P_z=0$.

We can fit the two- and three-point correlators to the energy-eigenstate expansion as
\begin{equation} \label{eq:2pt fit form}
     C_\pi^{\text{2pt}}(P_z;t) = |A_{\pi,0}|^2 e^{-E_{\pi,0}t} +
                                 |A_{\pi,1}|^2 e^{-E_{\pi,1}t} + \ldots .
\end{equation}
and
\begin{multline} \label{eq:3pt-fit-form}
    \\C_\pi^\text{3pt}(z,P_z;t_{\text{sep}},t)=|A_{\pi,0}|^2 \langle 0|O_{g,tt}|0\rangle e^{-E_{\pi,0}t_{\text{sep}}} \\
   + |A_{\pi,0}| |A_{\pi,1}| \langle 0|O_{g,tt}|1\rangle e^{-E_{\pi,0}(t_{\text{sep}}-t) }e^{-E_{\pi,1}t} \\
  +  |A_{\pi,0}| |A_{\pi,1}| \langle 1 |O_{g,tt}|0\rangle e^{-E_{\pi,1}(t_{\text{sep}}-t) }e^{-E_{\pi,0}t} \\
   +  |A_{\pi,1}|^2 \langle 1|O_{g,tt}|1\rangle e^{-E_{\pi,1}t_{\text{sep}}} + \ldots \\
\end{multline}
The ground (first excited) state amplitudes and energies, $A_{\pi,0}$, $E_{\pi,0}$, $(A_{\pi,1}$, $E_{\pi,1})$ are obtained from the two-state fits of the two point correlators.
$\langle 0|O_{g,tt}|0 \rangle$, $\langle 0|O_{g,tt}|1\rangle = \langle 1|O_{g,tt}|0\rangle$, and $\langle 1|O_{g,tt}|1\rangle$ are ground- and excited-state matrix elements, which are extracted from the two-state simultaneous fits or the two-sim fits to the three-point correlator using various $t_{\text{sep}}$.

\begin{figure*}[htbp]
\centering
\includegraphics[width=0.8\textwidth]{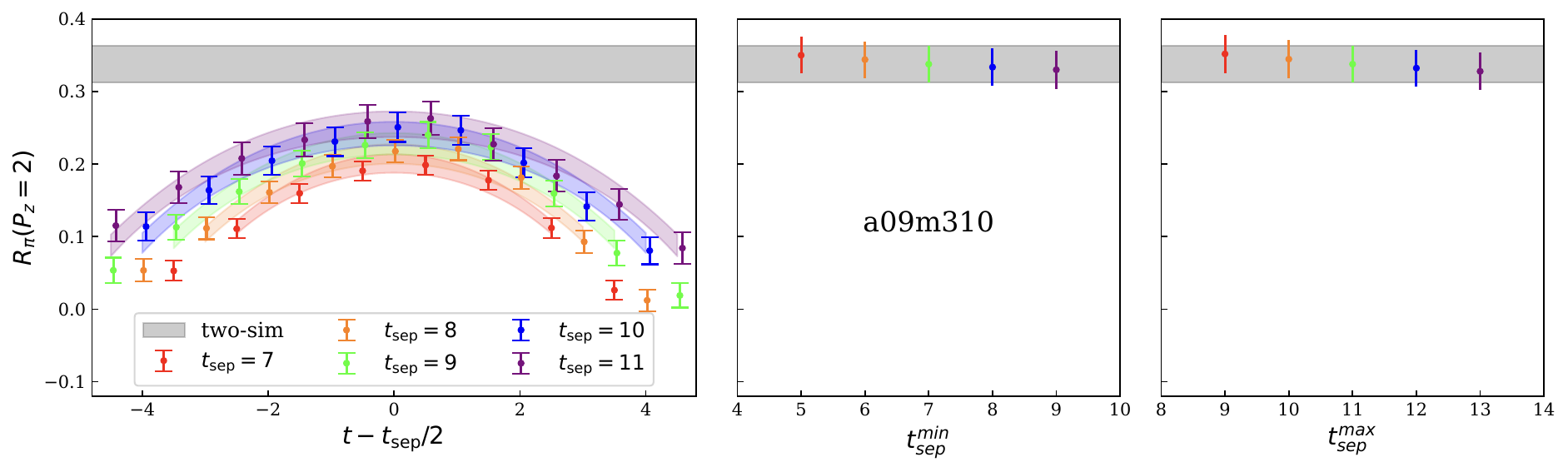}
\includegraphics[width=0.8\textwidth]{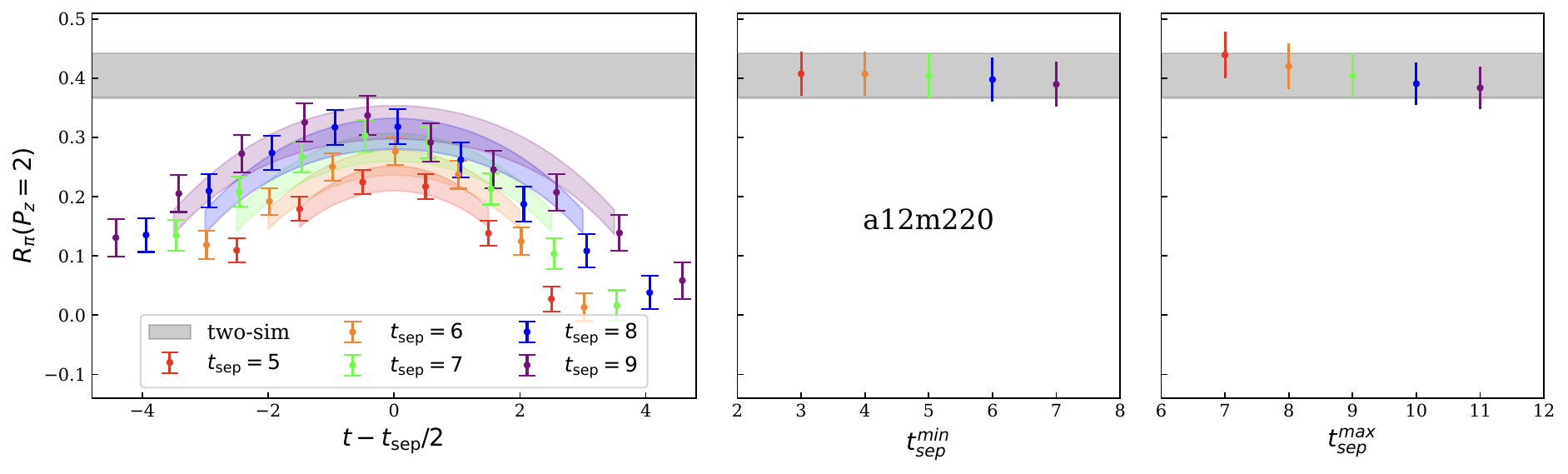}
\includegraphics[width=0.8\textwidth]{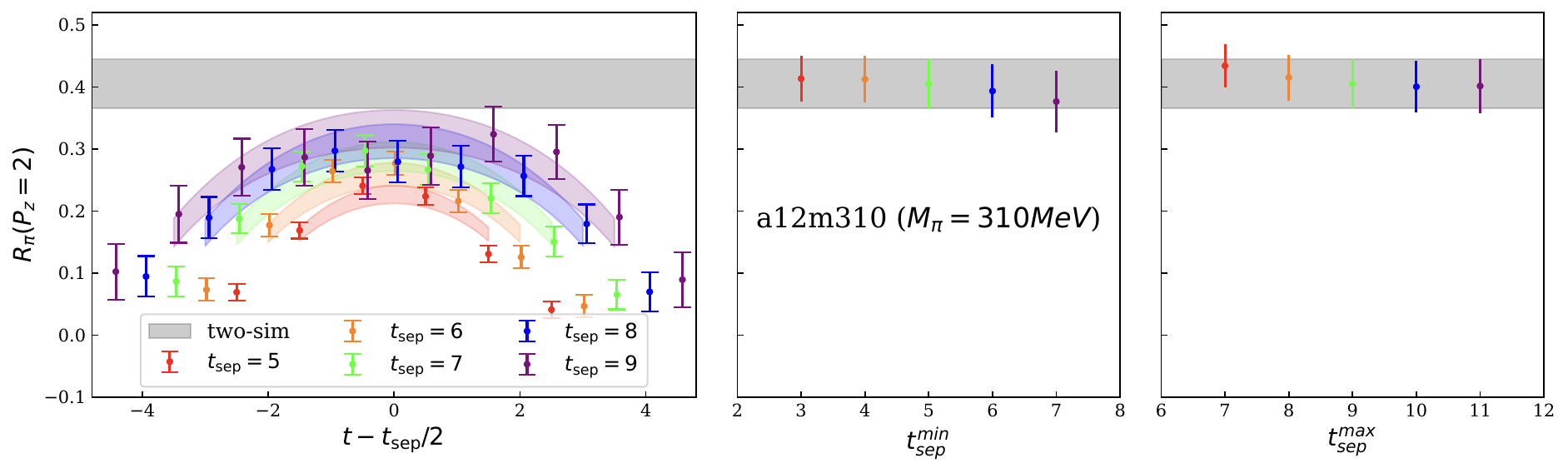}
\includegraphics[width=0.8\textwidth]{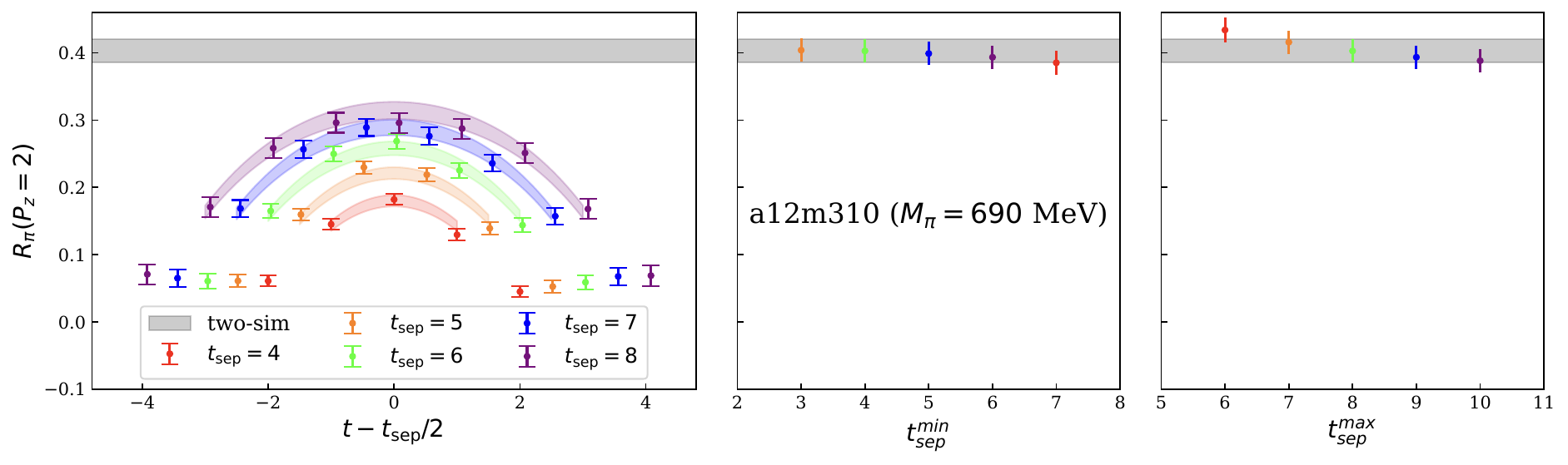}
\includegraphics[width=0.8\textwidth]{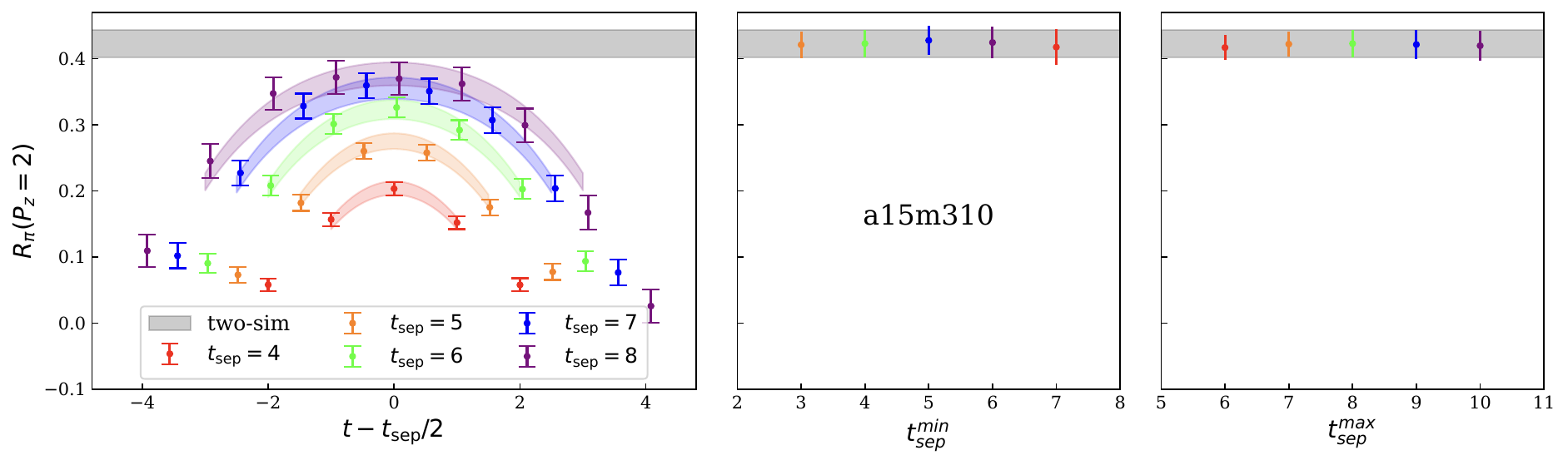}
\caption{
Example ratio plots (left column) and summary plots for the two-sim fits (last 2 columns) of the meson correlators at pion masses $M_\pi \in \{310, 220, 310, 690, 310\}$~MeV from the a09m310, a12m220, both a12m310, and a15m310 ensembles.
The gray band shown on each plot is the extracted ground-state matrix element from the two-sim fit that we use as our best value.
From left to right, the columns are:
the ratio of the three-point to two-point correlators with the reconstructed fit bands from the two-sim fit using the final $t_\text{sep}$ inputs, shown as functions of $t-t_\text{sep}/2$,
%the one-state fit results for the three-point correlators at different $t_\text{sep}$ values, 
the two-sim fit results using
$t_\text{sep} \in[t_\text{sep}^\text{min},t_\text{sep}^\text{max}]$
varying $t_\text{sep}^\text{min}$
and $t_\text{sep}^\text{max}$.
\label{fig:LRatio-fitcomp}
}
\end{figure*}

To visualize the reliability of the fitted matrix elements, we compare the fits to the ratios of the two- and three-point correlators:
\begin{equation} \label{eq:ratio}
    R_\pi(P_z;t_{\text{sep}},t) = \frac{ C_\pi^\text{3pt}(P_z;t_{\text{sep}},t)}{ C_\pi^\text{2pt}(P_z;t_{\text{sep}})}
\end{equation}
If the excited-state contamination were small, the ratios would approach the ground-state matrix element. 
We carry out the fitting procedure described above for all values of $P_z$ on the five ensembles given in Table~\ref{tab:LatticeParameters}.
Figure~\ref{fig:LRatio-fitcomp} shows examples of two-sim fits used to extract the ground-state matrix elements at $P_z=2 \times \frac{2 \pi}{aL}$ on our ensembles.
The leftmost column shows the ratios $R_\pi$ at different source-sink separations $t_{\text{sep}}$ (red to purple points), the reconstruction bands of the fits to the ratio plots (red to purple bands), and the fitted ground-state matrix element $\langle 0|O_{g,tt}|0 \rangle$ (gray band).
We observe that the $R_\pi$ data points have a tendency to increase with an increase in source-sink separation $t_{\text{sep}}$ and move upward toward the true ground-state matrix element.
Equation~\ref{eq:ratio} suggests that the ratios should be symmetric towards the source and sink (start and end of $t-t_{\text{sep}}/2$) as indicated by the reconstructions bands.
We see this trend clearly in the a09m310, a12m310 ($M_\pi=690$~MeV) and a15m310 ensembles.
This is also the case for the lower $t_{\text{sep}}$ data points, $t_{\text{sep}}\in\{5,6\}$ for the a12m310 ($M_\pi=310$~MeV) and a12m220 ensembles.
However, there seems to be some deviation in this trend towards the higher source-sink separations in the a12m310($M_\pi=310$~MeV) and a12m220 ensembles.
This deviation could be caused by statistical fluctuations due to increase in the signal-to-noise ratio at larger $t_{\text{sep}}$.
Even though the central value of the data points deviates, within two standard deviations (90\% confidence level) the ratio data points do display symmetry around $t-t_{\text{sep}}/2=0$.

We also study the dependence of source-sink separation choice in our simultaneous two-state fits to determine the pion ground-state matrix elements.
This enables us to see if our extracted ground-state matrix element is stable under various choices of $t_{\text{sep}}^{\text{min}}$ and $t_{\text{sep}}^{\text{max}}$.
The middle column of Fig.~\ref{fig:LRatio-fitcomp} shows the ground-state matrix elements as we vary $t_{\text{sep}}^{\text{min}}$, while keeping $t_{\text{sep}}^{\text{max}}$ constant at 11, 9, 9, 8 and 8 for a09m310, a12m220,
a12m310 ($M_\pi=310$~MeV), a12m310 ($M_\pi=690$~MeV) and a15m310 ensembles, respectively.
In each row, the gray bands in the middle and rightmost plots are obtained from the best choice of $t_{\text{sep}}$ range, as shown in the leftmost plots. 
The colored points show the ground-state matrix elements for different choices of $t_{\text{sep}}^{\text{min}}$.
The green point shows our final choice for the two-sim fits used in the leftmost column.
As the figure shows, the ground-state matrix elements converge and are, therefore, consistent as we vary the $t_{\text{sep}}^{\text{min}}$.

The rightmost column of Fig.~\ref{fig:LRatio-fitcomp} shows how the ground-state matrix elements change as we vary $t_{\text{sep}}^{\text{max}}$, while keeping $t_{\text{sep}}^{\text{min}}$ constant at 7, 5, 5, 4 and 4 from top to bottom in the plots in Fig.~\ref{fig:LRatio-fitcomp}.
Just as in the middle column, the gray band represents the ground-state matrix elements using the $t_{\text{sep}}$ obtained from the leftmost plot of each row.
We see the ground-state matrix element converge at higher $t_{\text{sep}}^{\text{max}}$ for all the ensembles and at lower $t_{\text{sep}}^{\text{max}}$ for the a09m310 and a015m310 ensemble.
However, we see a slight deviance from the true ground-state matrix element at lower $t_{\text{sep}}^{\text{max}}$ for a12m220 and a12m310 ($M_\pi=310$~MeV) ensembles.
This deviance is most significant for the lowest $t_{\text{sep}}^{\text{max}}$ shown for a12m310 ($M_\pi=690$~MeV) ensemble.
This is expected, since smaller $t_{\text{sep}}^{\text{max}}$ has more significant excited-state contribution.
However, this does not cause problems for our ground-state matrix element extractions, since the final $t_{\text{sep}}^{\text{max}}$ we use is the green point in the plot, and we observe that the ground-state matrix element is stable as one goes to larger $t_{\text{sep}}^{\text{max}}$.
Using the same process for other $P_z$ we are  able to deduce $t_{\text{sep}}$ ranges for the five ensembles.
Our final choices for $t_{\text{sep}}$ range used for the rest of this work is given in Table~\ref{tab:LatticeParameters}.

\begin{figure*}[htbp]
\centering
\includegraphics[width=0.32\textwidth]{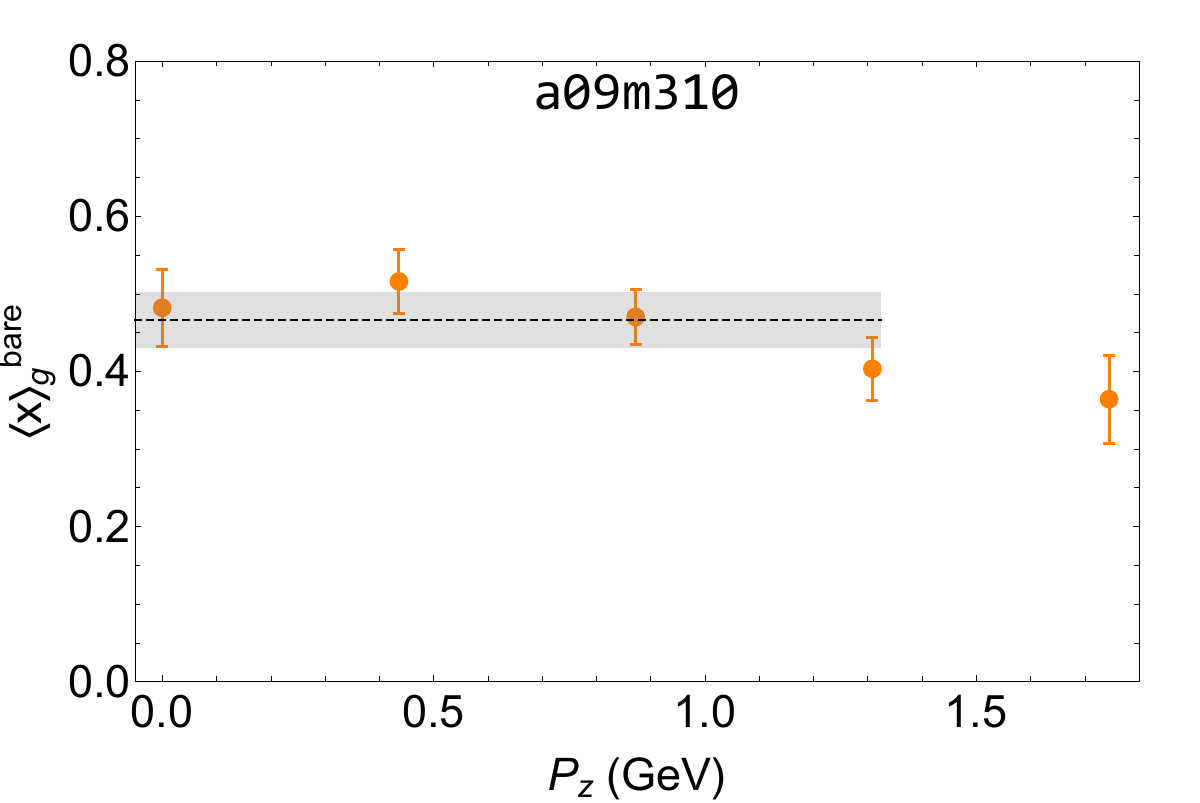}
\includegraphics[width=0.32\textwidth]{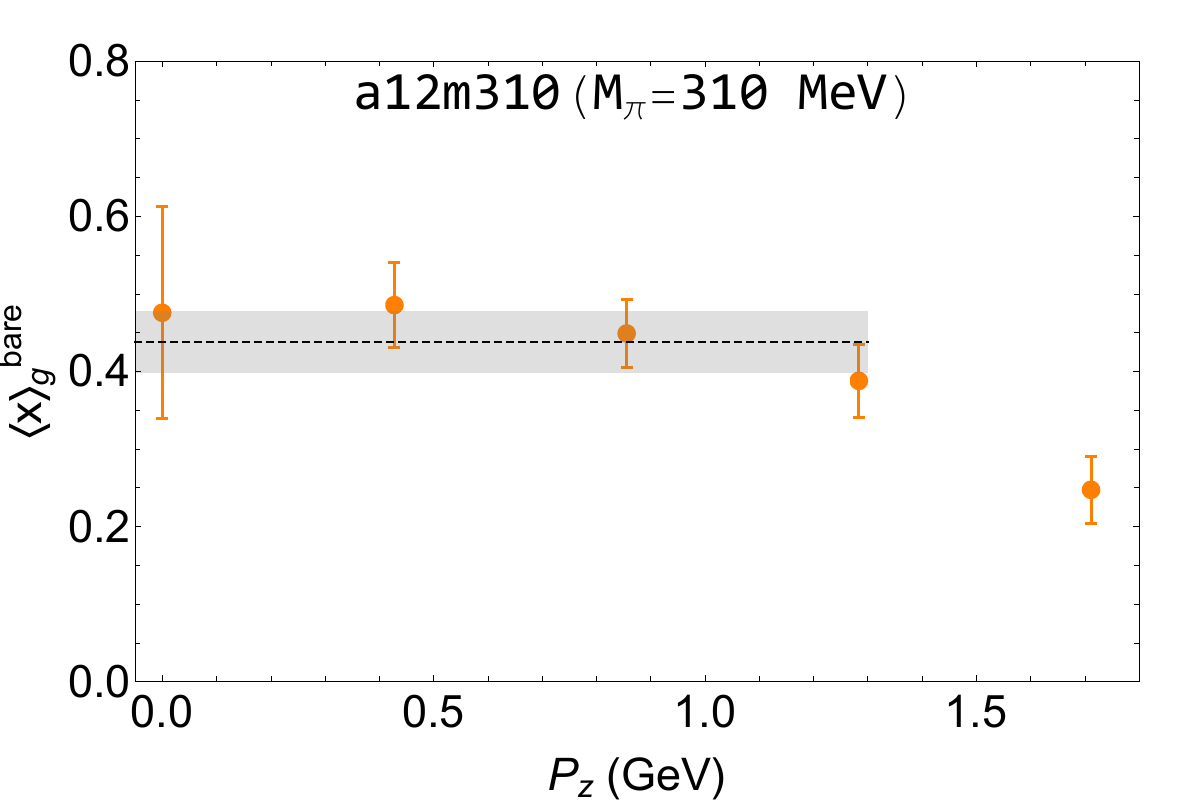}
\includegraphics[width=0.32\textwidth]{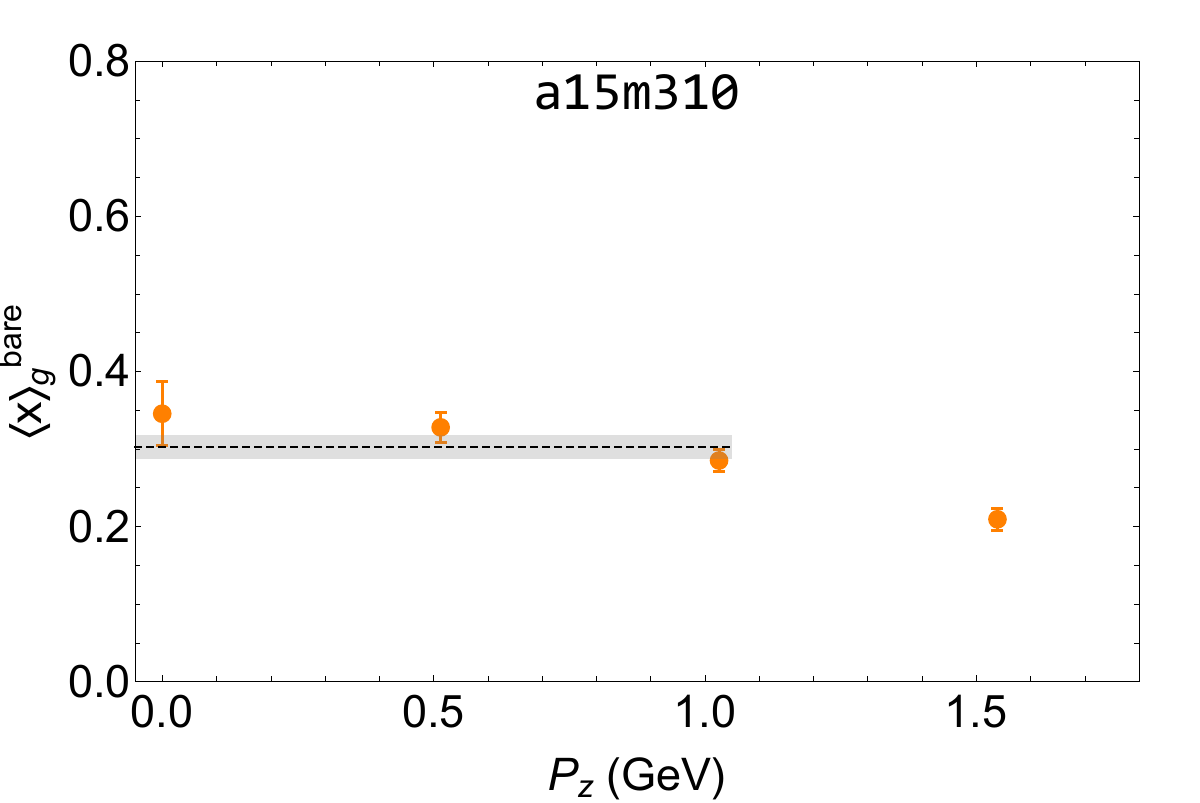}
\includegraphics[width=0.32\textwidth]{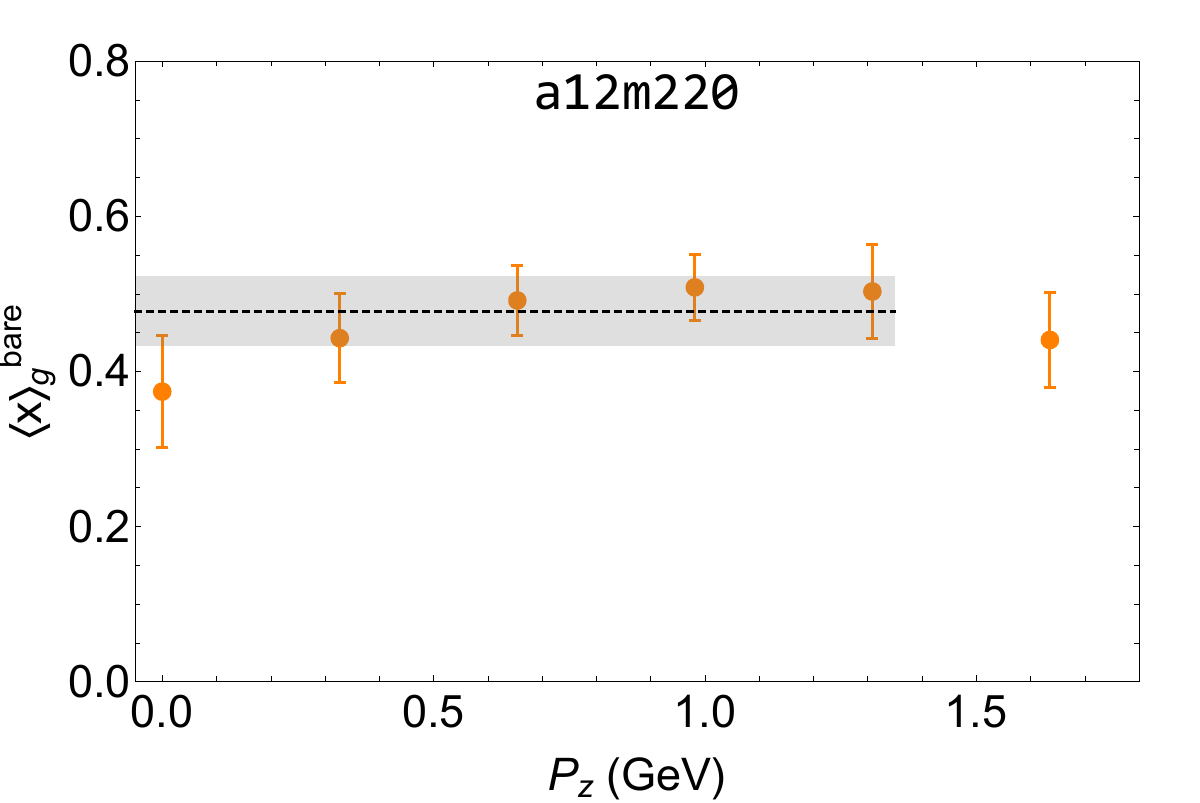}
\includegraphics[width=0.32\textwidth]{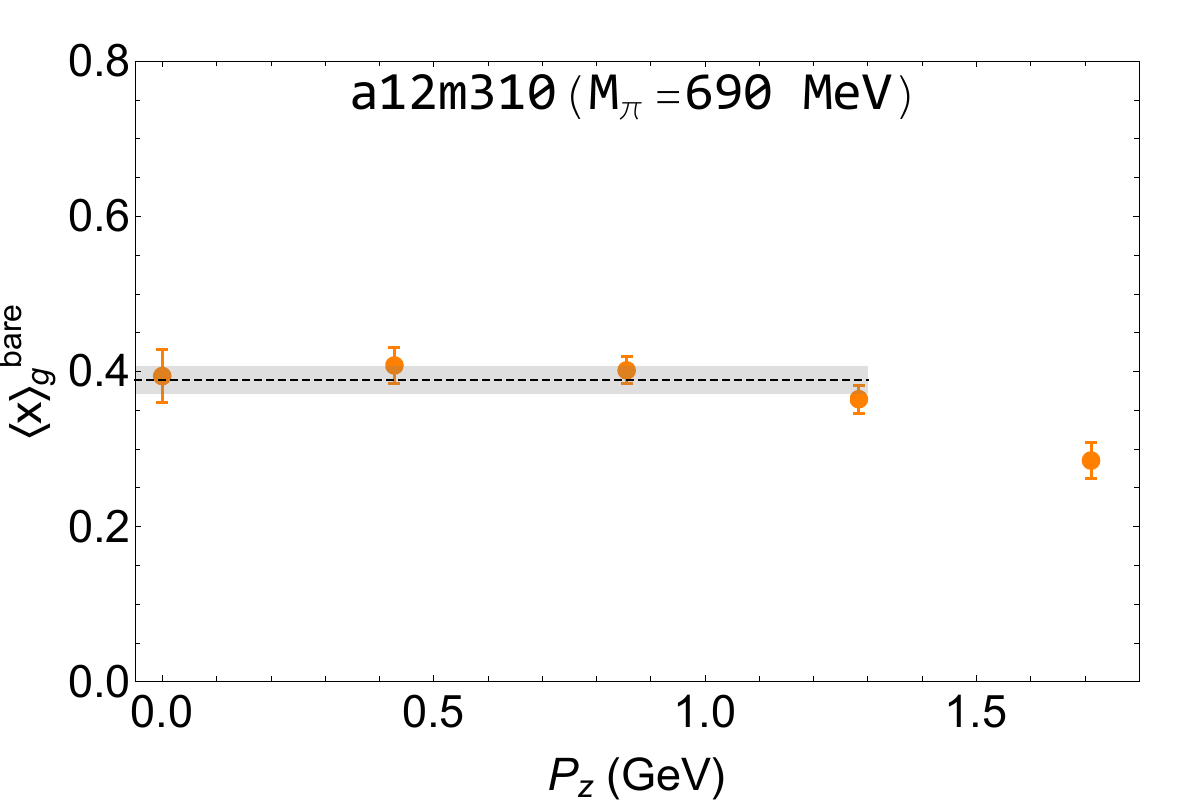}
\centering
\caption{
The bare gluon momentum fraction $\langle x \rangle_g^\text{bare}$ as a function of the momentum $P_z$ in GeV (orange points), along with weighted-average fitted pion gluon moment (grey band) for each ensemble.
The extent of the fit bands represents the data points used in the fit.}
\label{fig:BareME-fits}
\end{figure*}

To obtain the bare gluon moment matrix element $\langle x \rangle_g^\text{bare}$, we first need to multiply these matrix elements $\langle 0|O_{g,tt}|0\rangle$ obtained from the fit to the form in Eq.~\ref{eq:3pt-fit-form} at various momenta by the kinematic factor $\frac{4E_0}{3E_0^2+P_z^2}$.
These are shown as the orange data points in Fig.~\ref{fig:BareME-fits} as a function of momentum for the four ensembles used in this work.
We note that since our two-point correlators are momentum-smeared, our zero-momentum $\langle x \rangle_g^\text{bare}$ does not have the best signal-to-noise ratio.
Rather, the best data around found in the mid-momentum region of the plot.
We also note that the $\langle x \rangle_g^\text{bare}$ begin to deviate from constant at the largest momenta.
This is understandable, since as momentum increases, the discretization systematic grows like $P_z a$.
The effect is visible in our data: for a09m310, the data agree within two standard deviations, while the size of the deviation increases at coarser lattice spacing (see the top-row plots of Fig.~\ref{fig:BareME-fits}).

\begin{table*}[htbp!]
\centering
\begin{tabular}{|c|c|c|c|c|}
\hline
ensemble & $M_\pi^{\text{val}}$ (MeV) & $\langle x \rangle_g^{\text{bare}}$ & $\left( Z_{O_g}^{\overline{\text{MS}}} \right)^{-1}$ & $\langle x \rangle_g^{\overline{\text{MS}}}$ \\
\hline
a12m220 & 226.6(3) & 0.477(45) & 1.512(65) & $0.316(29)_\text{stat}(14)_\text{NPR}$ \\
\hline
a09m310 & 313.1(13) & 0.466(36) & 1.336(106) & $0.349(26)_\text{stat}(28)_\text{NPR}$ \\
\hline
\multirow{2}{*}{a12m310} & 309.0(11) & 0.438(40) & 1.512(65) & $0.290(25)_\text{stat}(13)_\text{NPR}$ \\
\cline{2-5}
& 684.1(6) & 0.389(18) & 1.512(65) & $0.257(11)_\text{stat}(11)_\text{NPR}$\\
\hline
a15m310 & 319.1(31) & 0.302(16) & 1.047(41) & $0.289(15)_\text{stat}(11)_\text{NPR}$ \\
\hline
a0m135 & 135 & -- & -- & $0.364(38)_{\text{stat}+\text{NPR}}(36)_\text{mixing}$ \\
\hline
\end{tabular}
\caption{
The bare gluon momentum fraction $\langle x \rangle_g^{\text{bare}}$, renormalization constant $\left( Z_{O_g}^{\overline{\text{MS}}} \right)^{-1}$, and renormalized gluon momentum fraction $\langle x \rangle_g^{\overline{\text{MS}}}$ for each ensemble used in this calculation.
We use the same NPR factors for the three different pion masses for the $a \approx 0.12$~fm ensembles, since the mass dependence of the factor was found to be weak.
In the final column, the first error is the statistical error from the matrix elements, and the second error is that from the NPR factor.
The last row gives the final extrapolated result for the momentum fraction.
In this result, the statistical and NPR error are combined as a result of the extrapolation process, making the first error, and the second is an estimate of 10\% mixing between the quark and gluon terms in the operator.
\label{tab:MEs}}
\end{table*}

To take advantage of our results at multiple momenta, we take the ensemble weighted average for the bare gluon momentum  $\langle x \rangle_g^\text{bare}$ all the calculated momentum data except for the largest one. 
These ensemble averages are represented by the gray bands in Fig.~\ref{fig:BareME-fits}.
The extent of the gray bands shows which $P_z$ we used in the weighted average.
The largest $\chi^2/\text{dof}$ for the weighted average is $2.2(16)$ for the a15m310 ensemble, due to the small number of points used.
The rest of the ensembles have $\chi^2/\text{dof} \in [0.7,1.3]$.
The ensemble average gluon bare momentum fractions are listed in Table~\ref{tab:MEs}

%%%%%%%%%%%%%%%%%%%%%%%%%%%%%%%%%%%%%%%%%%%%%%%%%%%%%%%%%%%%%%%%%%%%%%%%%%%%%%%%
\section{Results and Discussion}\label{sec:Results}

The leading moment of the pion gluon distribution requires renormalization of the bare operators.
The renormalization of the gluon operator can be carried out nonperturbatively on the lattice in the regularization-independent momentum-subtraction (RI/MOM) scheme and then converted to the modified minimal-subtraction scheme $\overline{\text{MS}}$.
The gluon moment renormalization in this work follows the same procedure as described in Refs.~\cite{Yang:2018bft,Shanahan:2018pib,Fan:2022qve}.
The renormalized gluon momentum fraction in the $\overline{\text{MS}}$ scheme can be written as
\begin{multline}
\label{eq:x_renorm}
\langle x \rangle_g^{\overline{\text{MS}}}
 = Z^{\overline{\text{MS}}}_{O_g}\left(\mu^2,\mu_R^2\right) \langle x \rangle_g^\text{bare} \\
 = R^{\overline{\text{MS}}}\left(\mu^2,\mu_R^2\right) Z^\text{RI}_{O_g}\left(\mu_R^2\right) \langle x \rangle_g^\text{bare},
\end{multline}
where the $Z_{O_g}$ are renormalization constants, and $\mu$ and $\mu_R$ are the renormalization scale used in the $\overline{\text{MS}}$ and RI/MOM schemes, respectively.
The one-loop perturbative matching ratio $R^{\overline{\text{MS}}}(\mu,\mu_R)$ is derived in Ref.~\cite{Yang:2016xsb} to convert the two renormalization schemes
\begin{multline}
\label{eq:matching_R}
R^{\overline{\text{MS}}}\left(\mu^2,\mu_R^2\right) =
    1 - \frac{g^2N_f}{16\pi^2} \left[\frac{2}{3}\ln\left(\frac{\mu^2}{\mu_R^2}\right) + \frac{10}{9}\right] \\
    - \frac{g^2N_c}{16\pi^2}\left(\frac{4}{3} - 2\xi + \frac{\xi^2}{4}\right),
\end{multline}
where $N_f=4$ and $N_c=3$ are the number of flavors and colors, respectively, $\xi=0$ selects Landau gauge, $g^2=4\pi\alpha_s(\mu)$ is the coupling strength (with $\alpha_s(\mu)$ the coupling constant with $\mu=2$~GeV).

However, naive attempts to calculate $Z_{O_g}^{\overline{\text{MS}}}$ proved difficult due to signal-to-noise ratios under 100\%, especially for the ensembles with finer lattice spacing~\cite{Yang:2018bft,Fan:2022qve}.
In order to improve the signal, we employ a technique developed in $\chi$QCD collaboration~\cite{Yang:2018bft} called ``cluster-decomposition error reduction'' (CDER) to improve the signal.
The motivation for this method is that the correlators decay exponentially with the distance between operator insertions, so there is no point in integrating beyond the correlation length, since it would  only pick up noise.
This can be implemented by imposing cutoffs in the spatial integrals used to calculate the correlators as done in Ref.~\cite{Fan:2022qve} for the nucleon gluon moment on the same ensembles.
For this work, we will take the renormalization constants $\left(Z_{O_g}^{\overline{\text{MS}}}\right)^{-1}$ from Ref.~\cite{Fan:2022qve} which are shown in the fourth column in Table~\ref{tab:MEs}. 

%%%%%%%%%%%%%%%%%%%%%%%%%%%%
\subsection{Pion Gluon PDF}

\begin{figure*}[htbp]
\centering
\includegraphics[width=0.45\textwidth]{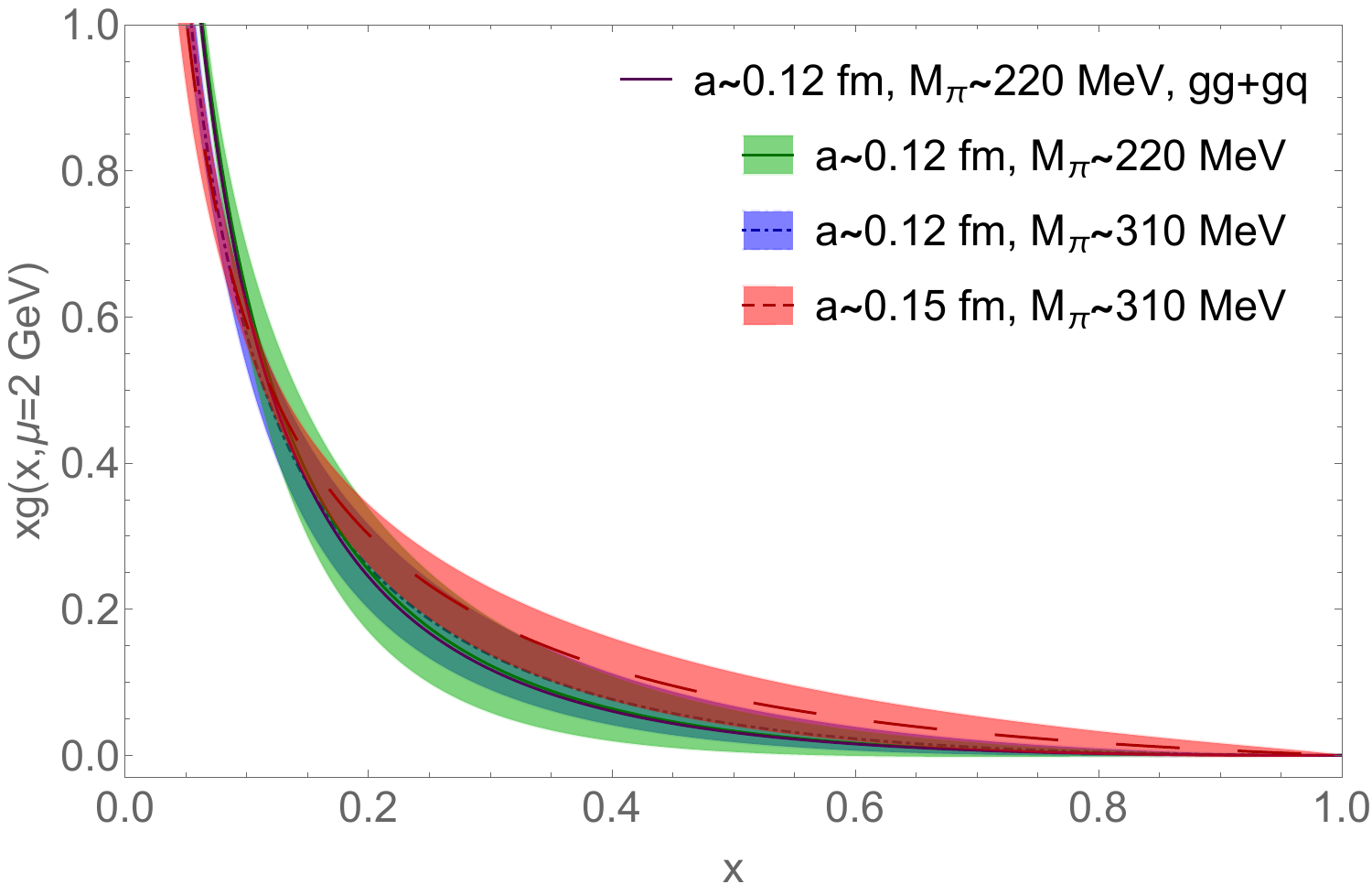}
\centering
\includegraphics[width=0.45\textwidth]{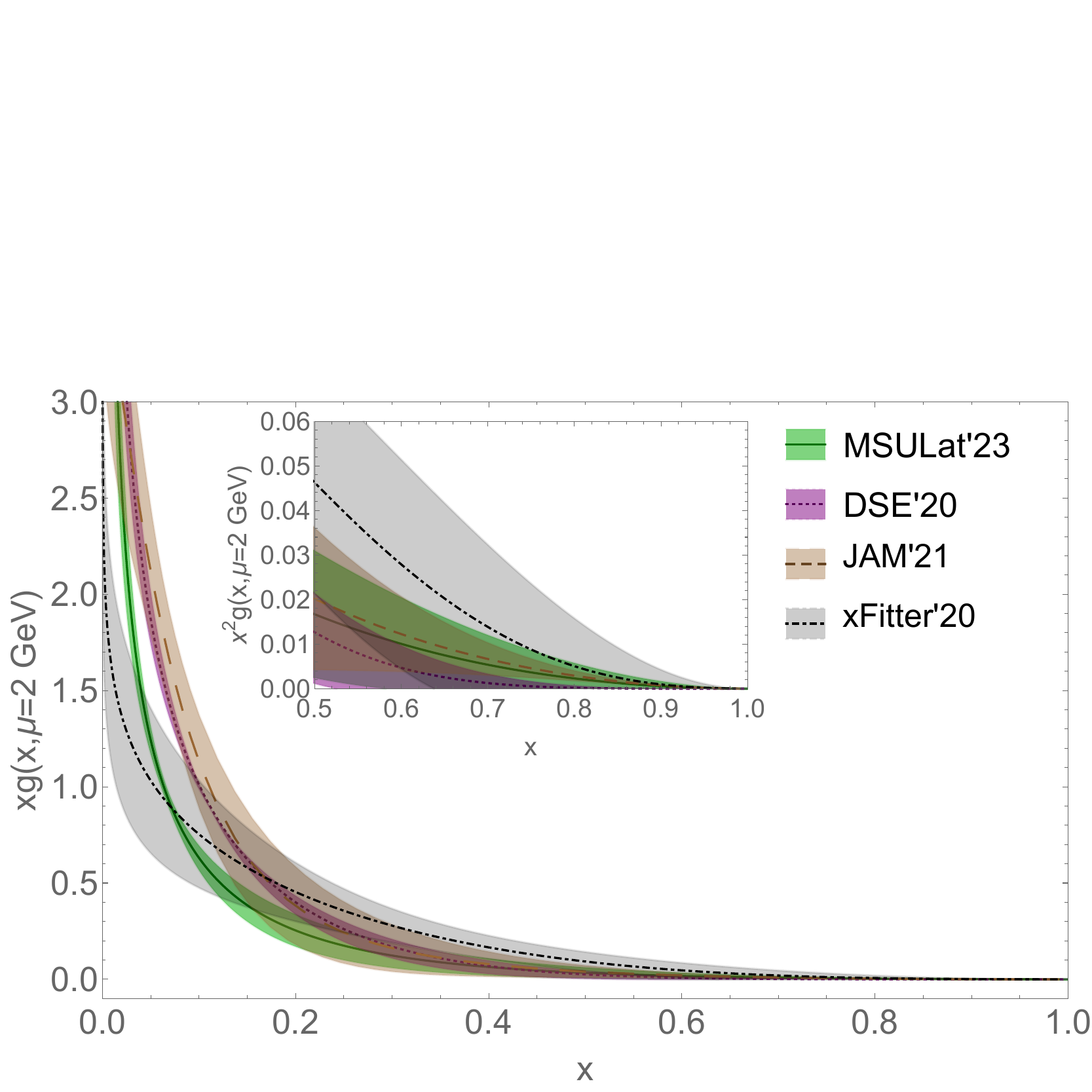}
\caption{
(Left) The pion gluon PDF $xg(x, \mu = 2\text{ GeV})$ as a function of $x$ for two lattice spacings $a \approx \{0.12, 0.15\}$~fm and pion masses $M_\pi = \{220, 310\}$~MeV.
These are calculated by multiplying the mean of the values of $\langle x \rangle_g$ from our results by the curves for $xg(x, \mu)/\langle x \rangle_g$ obtained in Ref.~\cite{Fan:2021bcr}.
Only the PDF errors and not the errors in $\langle x \rangle_g$ are represented here.
(Right) A comparison of the $a \approx 0.12$~fm $M_\pi \approx 220$~MeV $xg(x, \mu)$ result compared with the NLO pion gluon PDFs from xFitter'20~\cite{Novikov:2020snp} and JAM'21~\cite{Barry:2018ort,Cao:2021aci}, along with the DSE'20~\cite{Cui:2020tdf} at $\mu = 2$~GeV in the $\overline{\text{MS}}$ scheme.
The inset shows $x^2g(x, \mu)$ to highlight the agreements between the PDFs.
All the results are fairly consistent in the regions where $x > 0.2$.
}
\label{fig:PionPDF}
\end{figure*}

Using the renormalized gluon moment obtained from this work, we can now update the gluon PDF obtained by MSULat~\cite{Fan:2021bcr}.
In Ref.~\cite{Fan:2021bcr}, the authors used the ``pseudo-PDF'' method~\cite{Orginos:2017kos,Radyushkin:2018cvn,Balitsky:2019krf,Balitsky:2021cwr} to obtain the pion gluon PDF from overlap lattice ensembles with $a\approx 0.12$ and 0.15~fm and three pion masses $M_\pi\approx 220$, 310 and 690~MeV.
This work used the ratio renormalization scheme to avoid the difficulty of calculating the gluon renormalization factors, and therefore, could only obtain $xg(x, \mu)/\langle x \rangle_g$ as a function of $x$.
Since the pion $\langle x \rangle_g$ is not well known experimentally nor widely calculated on the lattice, the true gluon distribution of pion $g(x, \mu)$ was never determined on the lattice.
Using the $\overline{\text{MS}}$-scheme renormalized pion gluon moments from our a15m310 and a12m310 ensembles, we can extract $g(x, \mu)$ for the pion for the first time.
We multiply the binned $xg(x,\mu)/\langle x \rangle_g$ by the mean value of our $\langle x \rangle_g$ for each ensemble, not propagating errors, so the error bars on the lattice PDFs in Fig.~\ref{fig:PionPDF} are underestimated.

The $xg(x, \mu)$ pion PDFs are updated in Fig.~\ref{fig:PionPDF}.
We follow the convention commonly used in the global-fitting community by weighting the PDF with additional factor of $x$.
The left-hand side of Fig.~\ref{fig:PionPDF} shows pion gluon PDF at 310~MeV with lattice spacings of 0.12 and 0.15~fm.
The coarse lattice-spacing gluon PDF has slightly higher central values but is consistent with the one from finer lattice spacing over most $x$.
At 0.12-fm lattice spacing, we can study the pion-mass dependence of the gluon PDF at 310 and 220~MeV.
We find the heavier pion mass has slightly larger central value but remains consistent within statistical errors.
All three pion gluon PDFs are consistent with each other within the current statistical error.
We also look into potential mixing from the total quark contributions on the lightest pion result at 220~MeV, shown as a black solid line on left-hand side of Fig.~\ref{fig:PionPDF}, since there is no pion sea-quark distribution available from lattice QCD yet.
Taking the total quark distribution from global fits, we update the a12m220 gluon PDF to include gluon-in-quark (gq) contributions in the matching kernel, and we found the change to be very small.

We then compare our gluon PDF result at the lightest pion mass, 220~MeV, with the global fits~\cite{Barry:2018ort,Cao:2021aci,Novikov:2020snp} and phenomenological results from the Dyson-Schwinger equation (DSE)~\cite{Cui:2020tdf} on the right-hand side of Fig.~\ref{fig:PionPDF}.
The inset plot weights an additional factor of $x$ and provides a further zoomed-in view for a large-$x$ PDF comparison. 
From the right plot of Fig.~\ref{fig:PionPDF}, we see that our gluon PDF results are consistent with the results from JAM and the DSE for $x>0.2$.
The xFitter results are consistent with ours for $x > 0.15$ with slight tension around $0.3 < x < 0.375$.
The discrepancies in the small-$x$ region are likely due to lack of precision data in the small-$x$ region on the global-fit side and the lack of larger-momentum lattice matrix elements, which would provide better constraint of the results.
Overall, with the current accuracy in lattice calculation and global fits, there is reasonable agreement among them in the mid-to large-$x$ region.

%%%%%%%%%%%%%%%%%%%%%%%%%%%%
\subsection{Pion Gluon Moment in the Continuum-Physical Limit}

\begin{figure*}[htbp]
\centering
\includegraphics[width=0.45\textwidth]{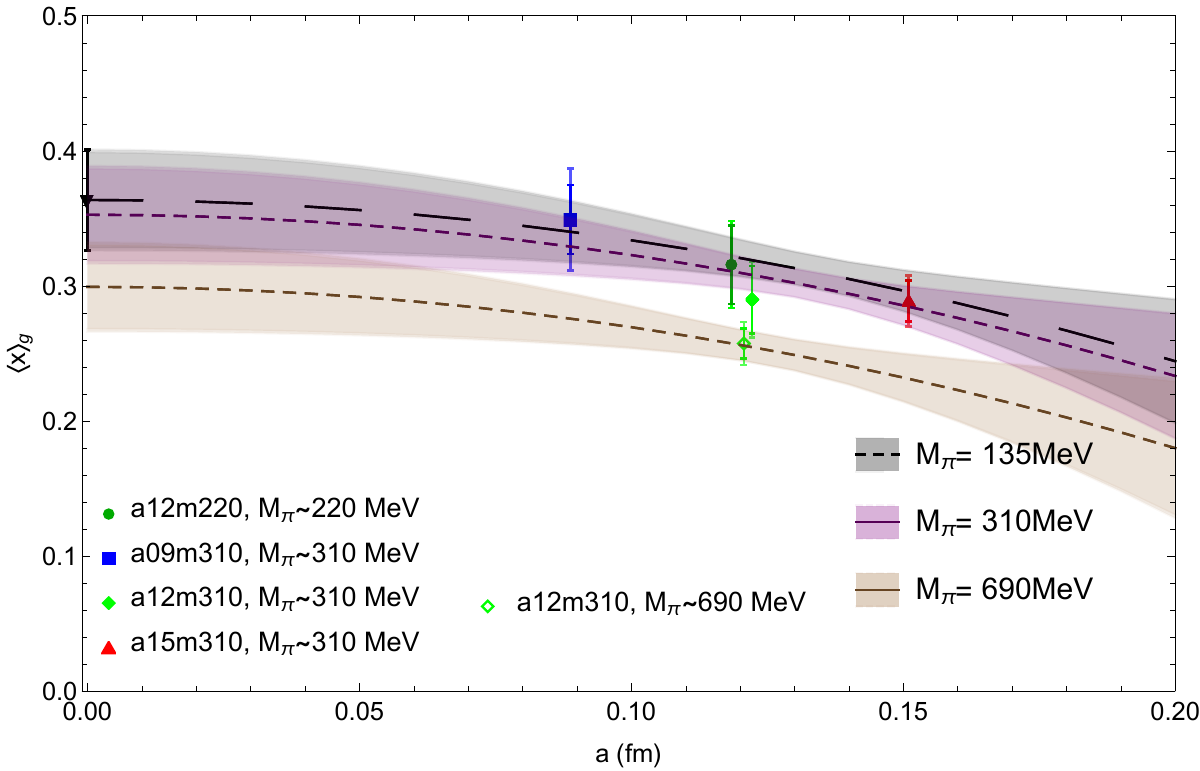}
\centering
\includegraphics[width=0.45\textwidth]{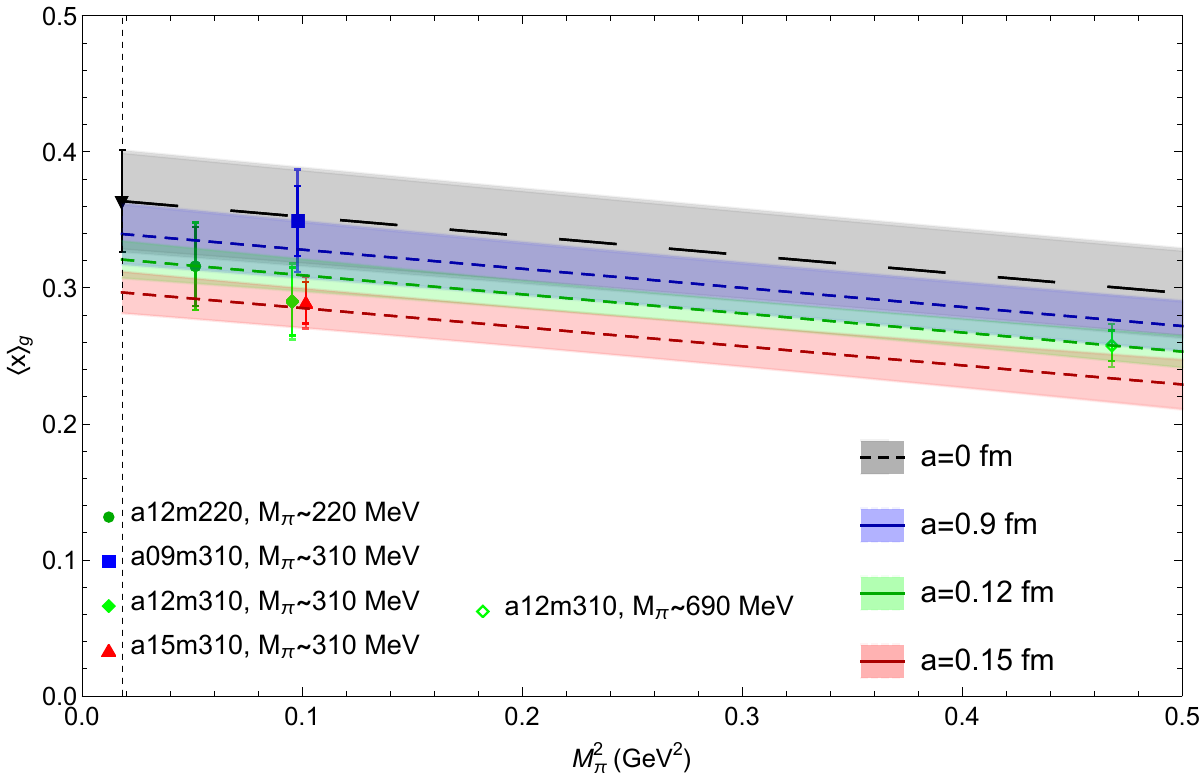}
\caption{
The renormalized gluon momentum fraction $\langle x \rangle_g$ from each ensemble and the physical-continuum extrapolation as functions of lattice spacing a (left) and pion mass $M_\pi^2$ (right)
Each data point in the plot has two errors: the darker inner bar indicates the statistical error, while the lighter outer bar includes combined errors from both the statistical and renormalization error.
The a12m310 data point for the lighter pion mass is shifted 0.0015~fm to the right on the $a$ plot for clarity.
The vertical dashed line in the right plot goes through $M_\pi^2 = (0.135\text{ GeV})^2$. The reconstructed fit bands at selected $M_\pi \in \{135, 310, 690\}$~MeV as functions of a and at selected $a \in \{0, 0.09, 0.12, 0.15\}$~fm as functions of $M_\pi$ are also shown in the left- and right-hand plots, respectively.}
\label{fig:Extraps}
\end{figure*}

With the renormalization constants and the bare momentum fractions $\langle x \rangle_g^{\text{bare}}$ obtained in Sec.~\ref{sec:lattice-details}, we summarize the renormalized $\langle x \rangle_g$ at the scale $\mu = 2$~GeV in ${\overline{\text{MS}}}$ scheme in Fig.~\ref{fig:Extraps}.
Each point in the figure has a darker and a lighter error bar, representing the statistical and systematic error from the gluon NPR factor, respectively.
Our renormalized $\langle x \rangle_g$ increase slightly with increasing pion mass but are consistent within total errors.
Similarly, we observe a small increase in the pion gluon moment toward the continuum limit;
however, the three 310-MeV data points are consistent within two standard deviations.
It will be interesting to study this further in the future with even higher statistics.

To obtain the continuum-physical limit pion gluon moment, we use a naive extrapolation function that is quadratic in $M_\pi$ and $a$
\begin{equation} \label{eq:phys-cont-ansatz}
    \langle x \rangle_g(M_\pi,a) = \langle x \rangle_g^{\text{cont}} + k_M(M_\pi^2 - (M^{\text{phys}}_\pi)^2) + k_a a^2.
\end{equation}
We perform the extrapolation fit with both statistical and NPR errors and obtain $\langle x \rangle_g^{\text{cont}} = 0.364(38)$ along with mass-dependence and lattice-spacing dependent fit parameters $k^{(\pi)}_M = -1.40(34) \times 10^{-4} \text{ GeV}^{-2}$ and  $k^{(\pi)}_a = -3.0(20)\text{ fm}^2$. 
Using the fit parameters, we can reconstruct the moment for $M_\pi \in \{135, 310, 690\}$~MeV ($a \in \{0, 0.09, 0.12, 0.15\}$~fm) as a function of $a$ ($M_\pi$),
as shown on the left (right) side of Fig.~\ref{fig:Extraps};
the fit well describes our measured data. 
Similar to the data points, each band has a statistical and total error that includes the NPR uncertainties;
however, the secondary error bands are too close to the statistical ones to be seen.

The left-hand side of Fig.~\ref{fig:Extraps} shows the lattice-spacing dependence of the pion gluon moments at fixed pion mass.
Overall, we see a trend towards a slightly higher pion gluon moment as the pion mass decreases.
The $M_\pi\approx 310$~MeV band has consistent pion gluon moment over the range $a \in [0,0.2]$~fm with the one extrapolated to the physical pion mass. 
The heavy pion mass $M_\pi\approx 690$~MeV shows the strongest deviation near $a\approx 0.12$~fm, but the band is within one sigma at $a = 0$.

On the right-hand side of Fig.~\ref{fig:Extraps}, we plot the pion-mass dependence at each lattice spacing used in this work, 0.15, 0.12 and 0.09~fm (red, green and blue, respectively), alongside the extrapolated band at the continuum limit, $a=0$ (grey band). 
These bands move upward approaching the continuum limit.
The coarsest lattice spacing used in this work has the strongest deviation from the continuum-limit results observed.
The $a=0.12$~fm ($a=0.09$~fm) band is consistent with the $a=0$ continuum band within two (one) standard deviation. 
Compared to the nucleon gluon moment, studied on similar lattice ensembles in Ref.~\cite{Fan:2022qve}, we see in the pion that the momentum-fraction dependence on the lattice spacing is slightly larger but still within two sigma of zero, while the pion-mass dependence is slightly larger and not consistent with zero.

There is an additional systematic error that we have not taken into account yet: gluon-quark mixing.
The bare gluon operator from Eq.~\ref{eq:gluon_op} can mix with singlet quark operators  through the renormalized gluon operator via $O_g = Z_{gg}O_g^\text{bare} + Z_{gq}\sum_{i=u,d,s}O_{q,i}^\text{bare}$.
An ETMC study~\cite{ExtendedTwistedMass:2021rdx} found the mixing to contribute as much as 20\% to the renormalized gluon operator in their study, while their earlier $N_f=2$ work~\cite{Alexandrou:2016ekb}, and an MIT study~\cite{Hackett:2023nkr} found the contribution to be no more than 10\%.
We choose to estimate a 10\% systematic error coming from the quark mixing, since we and MIT both use clover fermion action.
Therefore, our final results for pion gluon moment is $\langle x \rangle_g = 0.364(38)_{\text{stat}+\text{NPR}}(36)_\text{mixing}$ in the continuum-physical limit.

\begin{figure}[htbp]
\centering
\includegraphics[width=0.45\textwidth]{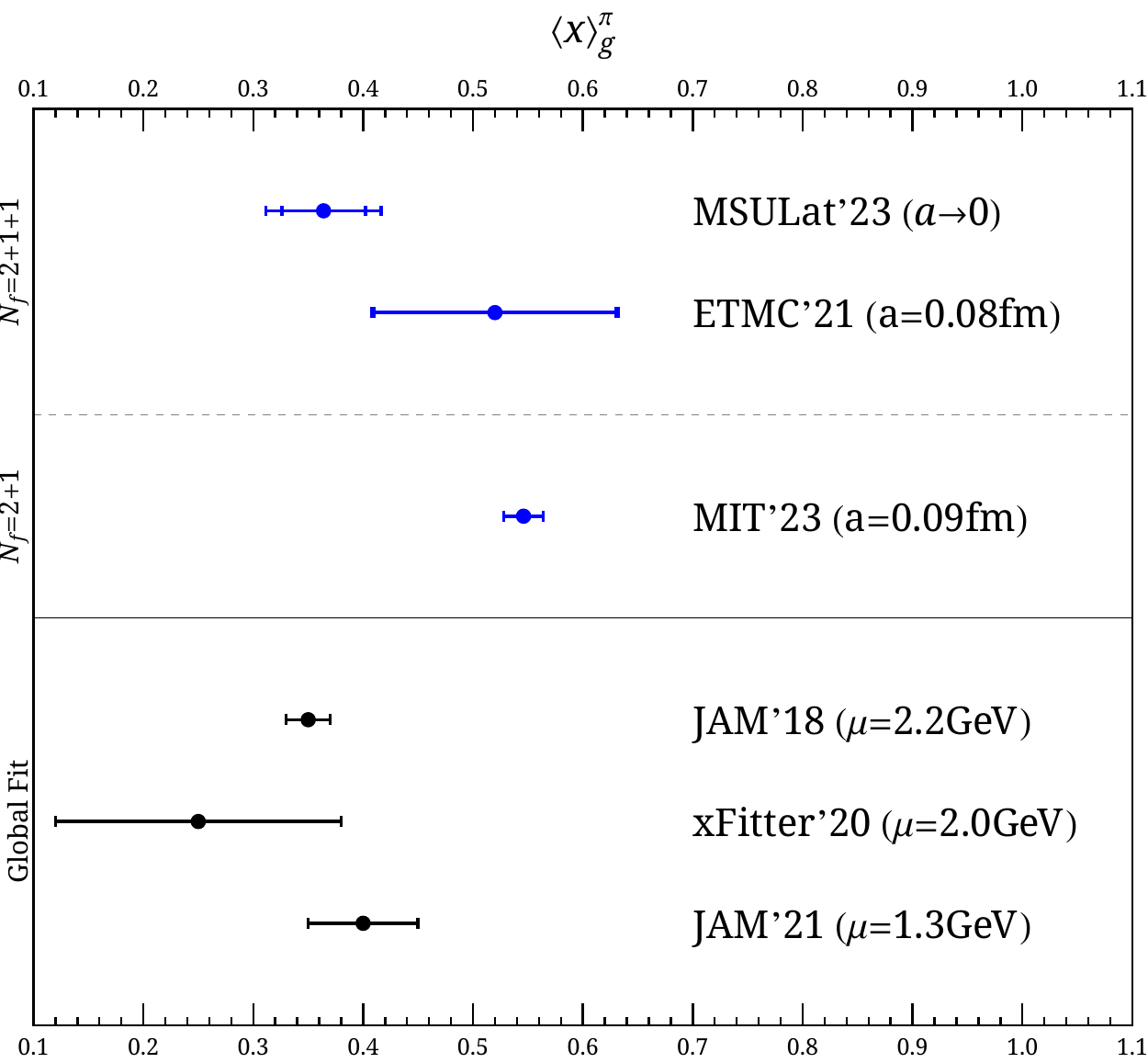}
\caption{
Comparison of lattice-QCD and global-fit determinations of the gluon moments of the pion.
The lattice-QCD results are given at $\mu=2$~GeV in $\overline{\text{MS}}$ scheme, and the global-fit results near 2~GeV as shown.
Lattice results are only shown if calculated at, near or extrapolated to physical pion mass.
These are
this work (MSULat'23),
ETMC'21~\cite{ExtendedTwistedMass:2021rdx} with $N_f=2+1+1$, and
MIT'23~\cite{Hackett:2023nkr} with $N_f=2+1$,
The global-fit results are
JAM'18~\cite{Barry:2018ort} at $\mu=2.2$~GeV
xFitter'20~\cite{Novikov:2020snp} at 2.0~GeV
and
JAM'21~\cite{Barry:2021osv} at 1.3~GeV.
Some of these numbers include systematic errors and some do not;
for those that do include them, the inner error bar is statistical only and outer error includes the estimated systematic errors.
We refer readers to Table~\ref{tab:latticemoments} for more details on the differences in the errors. 
Our lattice result (labeled MSULat'23) is consistent with the gluon momentum fraction obtained from the global fits.
}
\label{fig:xgpi-comp}
\end{figure}

In Fig.~\ref{fig:xgpi-comp}, we summarize the pion gluon moments obtained from lattice-QCD calculations done near or extrapolated to physical pion mass,  renormalized at 2~GeV in $\overline{\text{MS}}$ scheme, and from global fits by taking the first moment of their gluon PDF. 
Our continuum-limit result is consistent with another $N_f=2+1+1$ study by ETMC~\cite{ExtendedTwistedMass:2021rdx} at a single lattice spacing $a\approx 0.08$~fm, but is a few standard deviations away from the $N_f=2+1$ study by MIT group~\cite{Hackett:2023nkr}.
While it is far less accessible to perform such analyses for the pion as compared to the nucleon, recently, progress has been made towards a global QCD analysis yielding pion PDFs by the JAM~\cite{Barry:2021osv,Barry:2018ort} and xFitter~\cite{Novikov:2020snp} collaborations.
The xFitter collaboration uses Drell-Yan (DY) and prompt photon production data and applies their nucleon-PDF fitting framework to pion PDFs.
They obtain results consistent with ours but with much larger error bar. 
In the latest JAM global fit, using leading neutron electroproduction data from HERA in addition to data from pion-nucleus DY lepton-pair production and imposing the momentum sum rule, infers the value of $\langle x_g \rangle$ to be 0.40(3) using next-to-leading order (NLO) and next-to-leading logarithm (NLL) double-Mellin resummation at $\mu=1.3$~GeV.
The moment from JAM should increase when run to 2-GeV scale but should remain consistent with our results. 
We can also compare with QCD models, such as gauge-invariant nonlocal chiral quark model (NL$\chi$QM)~\cite{Hutauruk:2023ccw}, which gives $\langle x_g \rangle$ around 0.6.
The Dyson-Schwinger equation (DSE) framework~\cite{Ding:2019lwe,Cui:2020tdf} gives $\langle x_g \rangle$ 0.41(2) at 2~GeV, consistent with our result.
MAP collaboration~\cite{Pasquini:2023aaf} uses a light-front model calculation of the pion PDFs to obtain $\langle x_g \rangle = 0.37(5)$ at 2~GeV.
Overall, our pion gluon moment has good agreement with the global fits and most recent QCD-model calculations.

%%%%%%%%%%%%%%%%%%%%%%%%%%%%%%%%%%%%%%%%%%%%%%%%%%%%%%%%%%%%%%%%%%%%%%%%%%%%%%%%
\section{Conclusion}\label{sec:Conclusion}

In this work, we reported the first continuum-limit lattice calculation of the pion gluon momentum fraction, extrapolated to physical pion mass.
We used three lattice spacings ($a\approx 0.09$, 0.12 and 0.15~fm) and three pion masses (220, 310 and 690~MeV) on $N_f = 2+1+1$ highly improved staggered quarks (HISQ) with clover fermions.
We used high-statistics pion two-point correlators with O(10$^5$)--O(10$^6$) measurements.
The ground-state matrix elements are extracted using two-state fits to multiple source-sink separations.
We studied the choice of source-sink separation and found the ground-state pion matrix elements remain stable.
We then nonperturbatively renormalized our bare matrix elements using RI/MOM scheme with signal-to-noise improvement from cluster-decomposition error reduction technique.
The final results for the gluon moment calculated at each ensemble are reported at 2~GeV in $\overline{\text{MS}}$ scheme as summarized in Table~\ref{tab:MEs}.

Using our pion gluon moments, we updated the pion gluon PDFs from an earlier study~\cite{Fan:2021bcr} using pseudo-PDF method.
The $x$-dependent method normalized the gluon PDF by its moment, and only when the gluon moment is known, one can retrieve the true gluon PDF.
We compared the gluon PDF from the a12m220, a12m310 and a15m310 ensembles, and found the pion-mass and lattice-spacing effects to be negligible within the statistical errors.
We also compared our PDFs with the JAM and xFitter global-fit results, and they are in good agreement within the reliable $x$ region $x>0.2$.

We then took the continuum-physical limit by extrapolating our gluon moment using a simple ansatz with linear dependence in the square of lattice spacing and pion mass. 
We found a slightly larger pion mass and lattice spacing dependence in the pion gluon moment than the nucleon gluon moment. 
Our pion gluon moment in the continuum-physical limit is consistent with the prior $N_f=2+1+1$ lattice study but has tension with a prior $N_f=2+1$ lattice study.
However, both prior works near physical pion mass were done at single lattice spacing;
the difference can be caused by different lattice actions going to the continuum limit differently or other systematics.
We found our gluon moment to be in good agreement with those obtained from global fits and most recent QCD model calculations.
It will be interesting for future calculations to include ensembles at the physical pion mass and finer lattice spacing with improved statistics, to continue to improve the lattice calculation.

%%%%%%%%%%%%%%%%%%%%%%%%%%%%%%%%%%%%%%%%%%%%%%%%%%%%%%%%%%%%%%%%%%%%%%%%%%%%%%%%
\section*{Acknowledgments}
We thank MILC Collaboration for sharing the lattices used to perform this study.
We thank Matthew Zeilbeck for his early involvement of this project.
The LQCD calculations were performed using the Chroma software suite~\cite{Edwards:2004sx}.
%Computing resources
This research used resources of the National Energy Research Scientific Computing Center, a DOE Office of Science User Facility supported by the Office of Science of the U.S. Department of Energy under Contract No. DE-AC02-05CH11231 through ERCAP;
facilities of the USQCD Collaboration, which are funded by the Office of Science of the U.S. Department of Energy,
and supported in part by Michigan State University through computational resources provided by the Institute for Cyber-Enabled Research (iCER).
% Grant funding for the project
The work of WG is supported by partially by MSU University Distinguished Fellowship and by U.S. Department of Energy, Office of Science, under grant DE-SC0024053 ``High Energy Physics Computing Traineeship for Lattice Gauge Theory''. 
The work of KH is supported by the Professional Assistant program at Honors College at MSU. 
The work of HL is  partially supported by the US National Science Foundation under grant PHY 1653405 ``CAREER: Constraining Parton Distribution Functions for New-Physics Searches'', grant PHY 2209424,  and by the Research Corporation for Science Advancement through the Cottrell Scholar Award.

%%%%%%%%%%%%%%%%%%%%%%%%%%%%%%%%%%%%%%%%%%%%%%

\end{document}